\documentstyle[aps,prl,multicol,fancyheadings,epsfig,amsbsy,amssymb,amstex]{revtex}

\def\bbbc{{\mathchoice {\setbox0=\hbox{$\displaystyle\rm C$}\hbox{\hbox
to0pt{\kern0.4\wd0\vrule height0.9\ht0\hss}\box0}}
{\setbox0=\hbox{$\textstyle\rm C$}\hbox{\hbox
to0pt{\kern0.4\wd0\vrule height0.9\ht0\hss}\box0}}
{\setbox0=\hbox{$\scriptstyle\rm C$}\hbox{\hbox
to0pt{\kern0.4\wd0\vrule height0.9\ht0\hss}\box0}}
{\setbox0=\hbox{$\scriptscriptstyle\rm C$}\hbox{\hbox
to0pt{\kern0.4\wd0\vrule height0.9\ht0\hss}\box0}}}}
\newcommand{\beq}{\begin{eqnarray}} 
\newcommand{\eeq}{\end{eqnarray}} 

\begin{document}
\title{Striped superconductors in the extended Hubbard model.}
\author{Ivar Martin, Gerardo Ortiz, A. V. Balatsky, and A. R. Bishop}
\address{Theoretical Division, Los Alamos National Laboratory, Los
Alamos, NM 87545}

\date{\today }

\maketitle

\begin{abstract}
We present a minimal model of a doped Mott insulator that
simultaneously supports antiferromagnetic stripes and $d$-wave
superconductivity.  We explore the implications for the global
phase diagram of the superconducting cuprates. At the 
unrestricted mean-field level,
the various phases of the cuprates, including weak and strong
pseudogap phases, and two different types of superconductivity
in the underdoped and the overdoped regimes, find a natural
interpretation. We argue that on the underdoped side, the
superconductor is intrinsically inhomogeneous  -- striped coexistence of
of superconductivity and magnetism -- and 
global phase coherence is achieved through Josephson-like coupling
of the superconducting stripes.  On the overdoped side, the state is 
overall homogeneous and the superconductivity is of the classical BCS
type.
\end{abstract}
\pacs{Pacs Numbers: XXXXXXXXX}

\vspace*{-0.4cm}
\begin{multicols}{2}

\columnseprule 0pt

\narrowtext
\vspace*{-0.5cm}

Experimental evidence increasingly suggests that 
microscopic inhomogeneous ``stripe'' states are ubiquitous in the doped
cuprates, as well as in other complex electronic 
materials \cite{tranquada}. Nanoscale stripe morphologies have been
inferred in YBCO and LSCO from neutron scattering and angle
resolved photoemission experiments. In the superconducting phase,
the stripes appear to coexist with superconductivity in a range of
dopings without destroying the global phase coherence.
The main issue that we
address in this Letter is the nature of the superconducting state in
the presence of stripes.

Inhomogeneous quantum states in non-superconducting lattice models are
no less common than they are in the experiments. Many lattice models
possessing antiferromagnetic (AF) ground states at half-filling, when 
doped away from that filling develop stripes at the unrestricted mean-field
(MF) level \cite{schulz,zaanen,poil,machida}. Exact solutions may lead to
fluctuations that introduce dynamics into the MF solutions,
but are expected to preserve 
the qualitative features. The
lower the spatial dimension the more important the quantum fluctuations
are and, sometimes, MF solutions do not reproduce the exact large
distance physics of the problem. However, often they pick up the
low lying manifold of excited states which becomes relevant at low
enough temperatures. Also,  3-dimensional coupling in real materials
helps to reduce the effect of fluctuations.

We consider here a minimal model with stripes to illustrate our conclusions.
We employ the 2-dimensional 
one-band Hubbard  Hamiltonian with  hopping $t$ and  on-site
repulsion $U$ \cite{schulz}. Superconductivity is introduced by
including the nearest neighbor attraction $V$, which produces pairing
predominantly in the $d$-wave channel close to half-filling
\cite{micnas}. The effective minimal Hamiltonian is thus
\beq\label{H}
H = -t \sum_{\langle ij \rangle \sigma}{c^\dagger_{i\sigma}
c^{\;}_{j\sigma}} +  U \sum_{i}{n_{i\uparrow} n_{i\downarrow}} + V
\sum_{\langle ij \rangle}{n_i n_j},
\eeq
where the operator $c^{\dagger}_{i\sigma}$  ($c^{\;}_{j\sigma}$)
creates  (annihilates) an electron with spin $\sigma$ on the lattice
site $i$, and $n_i = c^{\dagger}_{i\uparrow}c^{\;}_{i\uparrow} + 
c^{\dagger}_{i\downarrow}c^{\;}_{i\downarrow}$ represents the electron
density on site $i$.
For our computations, we use the unrestricted mean-field approximation 
to this Hamiltonian,
\beq\label{H_MF}
H_{MF} = -t \sum_{\langle ij \rangle \sigma}{c^\dagger_{i\sigma}
c^{\;}_{j\sigma}} + U \sum_{i}{n_{i\uparrow} \langle n_{i\downarrow}
\rangle + \langle n_{i\uparrow}\rangle n_{i\downarrow}} \nonumber\\ +
\sum_{\langle ij \rangle}{c^{\;}_{i\downarrow} c^{\;}_{j\uparrow}
\Delta_{ij}^* + {\rm H.c.}} \ ,
\eeq
where $\Delta_{ij} = V \langle c_{i\downarrow} c_{j\uparrow}\rangle$ is
the MF superconducting order parameter. The direct Hartree terms in $V$
are neglected since the magnitude of the effective  nearest neighbor
attraction is expected to be much smaller than the on-site repulsion
$U$.  Hence, it should not affect the diagonal part of the Hamiltonian,
which is responsible for the charge and spin order.   Therefore, the
effect of $V$ in our model is limited to the generation of
superconducting correlations. We do not address the very
important issue of the  microscopic origin of the attraction $V$. Our
goal is only to construct a minimal model that may help to qualitatively
understand the rich phase diagram of the cuprates.  Unless stated
otherwise, standard parameter values used are $U = 4t$ and $V = -0.9t$.
This choice of parameters allows us to clearly demonstrate our conclusions
regarding the inhomogeneous superconducting phase.

To self-consistently solve the MF equations we use an iterative 
scheme. This consists of two stages which are repeated until convergence
is achieved: 
(1) Diagonalization of the MF Hamiltonian, and 
(2) Update of the MF parameters.
An important feature of our approach is that all physical quantities
are allowed to vary from one lattice site to another, e.g., $\langle
n_{i\uparrow}\rangle \neq \langle n_{i\downarrow}\rangle$ and $\langle
n_{i\sigma}\rangle \neq \langle n_{i+\alpha \sigma}\rangle$.
Generically, we are seeking inhomogeneous solutions whose typical
correlation lengths $\xi$ involve several lattice spacings. Therefore,
it is important that the simulated supercell size $N_x \times N_y$
(with periodic boundary conditions) is such that $N_{x,y} > \xi_{x,y}$. 
The Hamiltonian in Eq. (\ref{H_MF}) can be rewritten in the matrix
form, $H_{MF} = {\bf c}^{\dagger} \widehat{H} {\bf c}$, with  ${\bf c}
= (c^{\;}_{1\uparrow}, c_{1\downarrow}^\dagger, c^{\;}_{2\uparrow},
c_{2\downarrow}^\dagger, \cdots)^T$, where $\widehat{H}$  is a $(2 N_x
N_y) \times (2 N_x N_y)$ hermitian matrix. By applying a unitary
transformation $\alpha$ ($\alpha^{-1} = \alpha^*$)
the Hamiltonian matrix can be  diagonalized as
$\widehat{H} = \alpha \widehat{D} \alpha^{-1}$, with  $D_{nm} =
\delta_{nm} E_n$. The Hamiltonian
can be diagonalized in the Bogoliubov  quasiparticles,
\beq
\gamma_n = \sum_m{\alpha^{-1}_{n m} c_{m}},
\eeq
with energies $E_n$. By reexpressing the original creation-annihilation
operators in terms of the Bogoliubov quasiparticles, one can recompute
the parameters of the MF Hamiltonian. For example,
\beq
\langle n_{i\uparrow} \rangle &=& \langle c_{i\uparrow}^{\dagger}
	c^{\;}_{i\uparrow}\rangle =  \sum_{nm}\langle{\alpha^*_{i\uparrow,
	n} \gamma_n^\dagger \alpha_{i\uparrow, m} \gamma_m\rangle}
	\nonumber\\ &=&\sum_n{|\alpha_{i\uparrow, n}|^2 } n_F(E_n), 
\eeq
where $n_F(E_n)$ is the Fermi-Dirac distribution function. Repeated
until the convergence, the iterations produce the spatial profiles of 
the self-consistent density and order parameter.  

\begin{figure}[htbp]
  \begin{center}
   \includegraphics[width = 3.0 in]{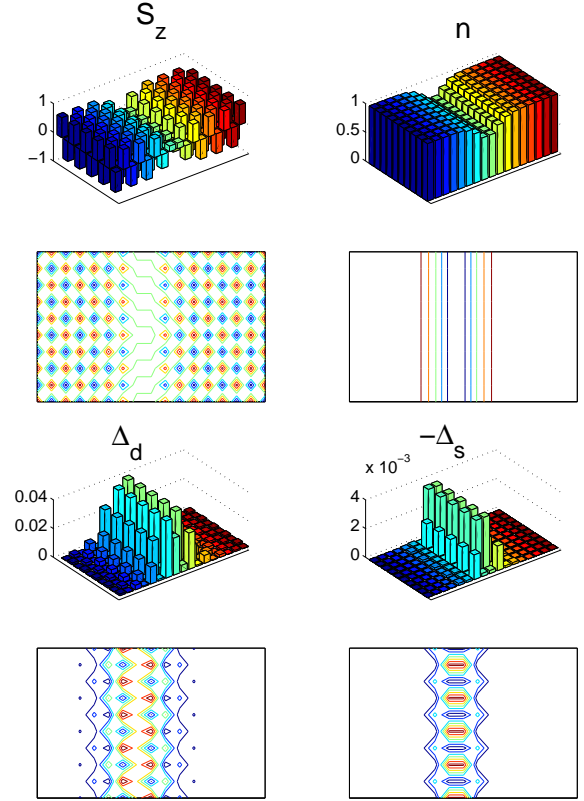}
\vspace{0.5cm}

\caption{Typical example of density and superconducting order 
parameter profiles in a
stripe state (here period 17).  The top two bar charts
represent the site-dependent spin and charge densities, respectively. 
The contour plots indicate the sites with low (blue) and high (red)
values of the corresponding densities. The bottom four plots show the
values of the superconducting order parameters, defined as 
$\Delta_i^{d(s^*)} = (\Delta_{i, right} + \Delta_{i, left}  \mp
\Delta_{i, up} \mp \Delta_{i, down})/4$  for $d$-wave (extended
s-wave)  order parameter on site $i$ ($U = 4t$, $V = -0.9t$). Different
choices of parameters lead to qualitatively similar patterns, with
stronger $U$ leading to a stronger AF order and more attractive $V$
causing the superconducting stripes to become wider and larger in
amplitude. The doping level is $5.9\%$.}

\label{fig:10x17}
\end{center}
\end{figure}

A typical zero-temperature MF inhomogeneous solution is shown in Fig.
\ref{fig:10x17}. In the lowest energy configuration, the spin density 
develops a soliton-like AF anti-phase domain boundary --- a stripe
--- at which the AF order parameter changes sign. At the domain
boundary, the electronic charge density is depleted. The width of
the domain wall, $\xi_{DW}$, decreases with increasing on-site
repulsion $U$. However, for values of $U$ that are not much larger
than the hopping $t$, the charge per unit length of the optimal (the
lowest energy) stripe remains the same and is close to unity near
half-filling. This result, first demonstrated in this simple model 
by Schultz \cite{schulz},
is a direct consequence of doping-dependent nesting in the Hubbard
model. The bond-centered stripes are favored relative to the 
site-centered ones, although the energy difference in our case is small
due to the smooth charge distribution.  
For a different band structure the exact relation between the
doping $x$ and inter-stripe distance, $L(x)$, may change; however, any
model whose ground state is AF at zero doping, is expected to have
AF stripes for a finite doping, with
incommensuration proportional to the doping, $1/L(x) \propto x$, near
half-filling.  For example, negative next-nearest neighbor hopping $t^\prime$
(relevant in the hole-doped cuprates\cite{ws}), 
modifies the stripe filling without
compromising the stripe phase stability relative to commensurate AF at the MF 
level\cite{ivar2}.  The stripe filling is a monotonically decreasing function
of the magnitude of $t^\prime$, with the magic filling 1/2 occurring when 
$t^\prime = -0.35 t$.  Experimentally, however, the value of the effective 
doping-dependent  $t^\prime$ is still not well defined.  
Among the consciences of the doping dependence of $t^\prime$ are
doping-dependent stripe filling, and a possibility of a transition 
from vertical
to diagonal stripes as a function of doping.
While we focus here
on the model case $t^\prime = 0$, our conclusions remain qualitatively the
same also in the presence of a moderate negative $|t^\prime|  \lesssim 0.5 t$.  

For large enough doping levels, such that 
$L(x) \lesssim \xi_{DW}$, the AF stripes begin to 
overlap.  In this regime the excitation spectrum is no longer fully gapped
and mobile carriers appear.
Further doping mostly changes the amplitude of the spin and charge density
waves, only slowly modifying the incommensuration, $1/L(x)$\cite{schulz}.
When the stripes are sufficiently close to melting, 
the AF aspect of the problem becomes unimportant,
and the superconductivity is of the conventional BCS type.

The superconducting order parameter $\Delta_{ij}^{d(s^*)}$ is maximal
on the stripes and is not smooth (even within the stripe) due to the
presence of the AF background. This happens since the order parameter
is sensitive to the spin density on sites $i$ and $j$. If $i$ belongs
to the spin-down sublattice and the neighbor $j$ is on the spin-up
sublattice, then the order parameter is large, and vice versa.  Notice,
that in addition to the dominant $d$-wave component, there is a small
extended $s$-wave component generated on the stripe, which can be
interpreted as a distortion of the $d$-wave at the level of about
$10\%$. This happens because a symmetry of the lattice has been broken. 
The superconducting stripe is pinned by the AF phase domain
boundary. Since the spin density on the domain boundary is small, a
natural interpretation is that the superconductivity is suppressed in
the regions of large AF order. Nevertheless, the width of a
superconducting stripe, $\xi_{SC}$, is determined not only by the 
width of the AF stripe, but also increases rapidly with increasing 
magnitude of the nearest neighbor attraction, $V$.  For our choice of
parameters, $\xi_{DW} \sim 4$ and $\xi_{SC} \sim 8$ lattice sites.
For dopings smaller than about $10\%$ (corresponding to $L(x) > 10$
lattice sites) the stripes have negligible overlap. In this regime, the 
amplitude of the superconducting order parameter on the stripes no longer 
depends upon the stripe-stripe separation. For higher doping levels, an overlap
between the superconducting order parameters on adjacent stripes
is established.

\begin{figure}[htbp]
  \begin{center}
    \includegraphics[width = 2.7 in]{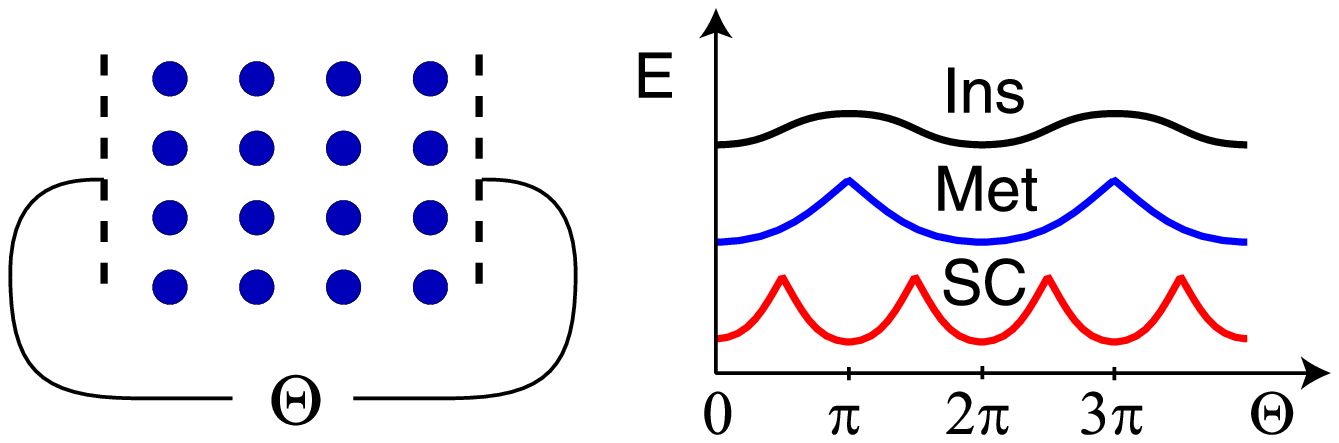}
    \includegraphics[width = 1.5 in]{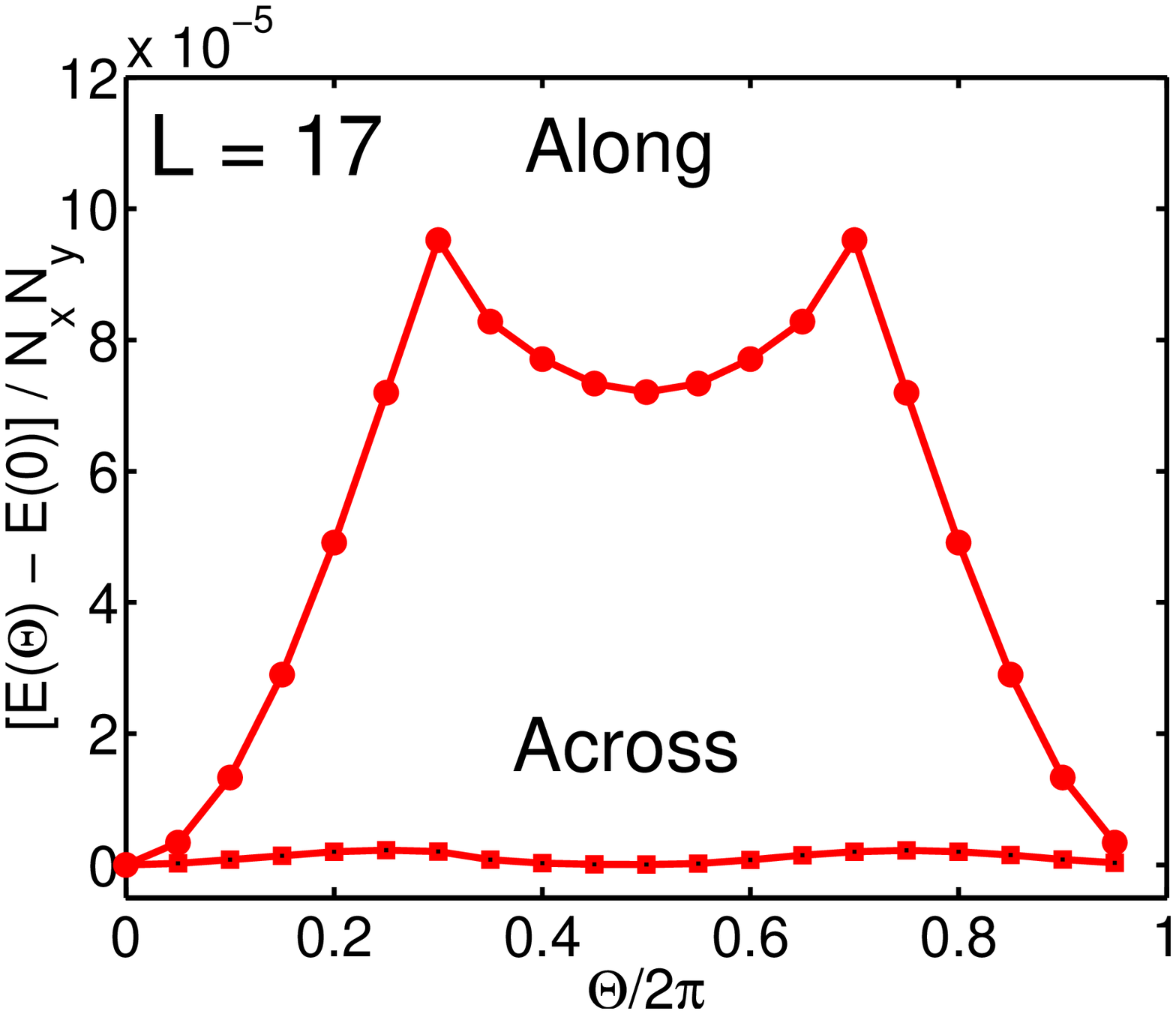}
    \includegraphics[width = 1.55 in]{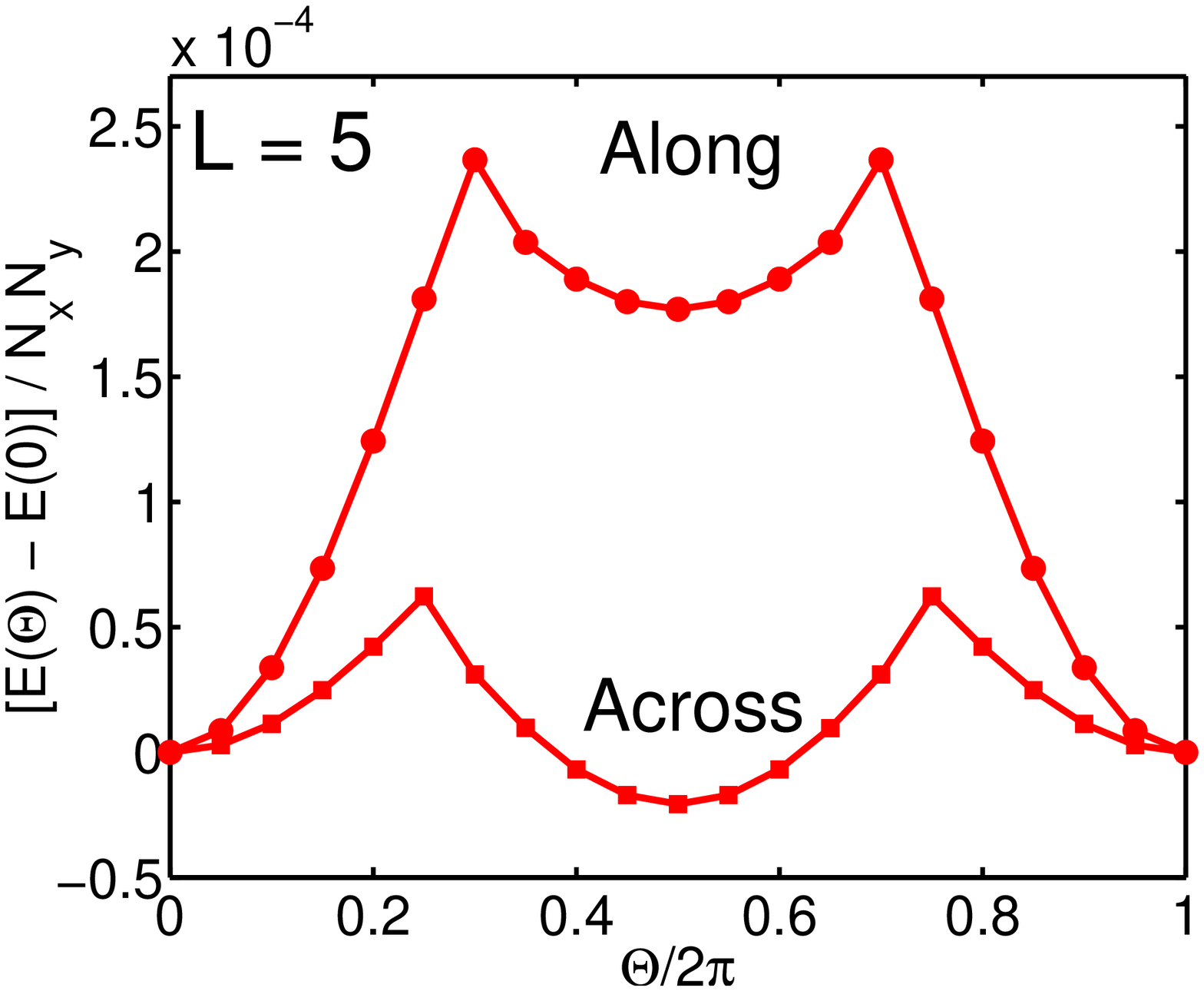}
\vspace{0.5cm}

\caption{ ({\bf A}) Schematics of typical energy spectra,
$E(\Theta)$. Insulating behavior is characterized by a smooth
curve of period $2\pi$,  with an amplitude that decays exponentially
with system size.  The  metallic and the superconducting states
typically have  patterns with cusps, with variations in $E(\Theta)$
that decrease as  a power of the system size in such a way that the
charge stiffness,  $D_c = \frac{L}{2} \partial^2 E/\partial \Theta^2$,
remains constant. The superconducting behavior is distinguished from
the metallic one  by the reduced periodicity, referred to as
anomalous flux quantization  (AFQ). There is a direct correspondence
between AFQ and the fact that the minimum flux that can penetrate a
superconducting system is one half of the flux quantum,  $\phi_0/2 =
\hbar c/2e$.  The exact period $\pi$ is only achieved in the 
thermodynamic limit; for a finite system, there is only a signature of
the reduced period. ({\bf B}) Calculated energy spectrum 
$E(\Theta)$ for a system
of size $10 \times 17$  (see Fig. \protect{\ref{fig:10x17}}).   {\em Along}
the stripes there is a pronounced AFQ signature, which implies that 
the system is  superconducting along
the stripes.  
The energies were computed at
a fixed chemical potential. To convert to the energy at a fixed
average number of particles, a density adjustment by the amount $-\mu
(n(\Theta) - n(0))$ has been made.
{\em Across} the stripes, the stiffness is  
very small, but has a period of $\pi$.
Hence the system appears to be a global superconductor, but with an extremely
small superfluid stiffness in the direction across the stripes.
({\bf C}) Calculated energy spectrum $E(\Theta)$ for a system
of size $10 \times 15$ with 3 stripes of  period 5.  Due to substantial overlap
between the superconducting stripes, this system has superconducting strengths 
that are comparable in both directions.
}
\label{fig:cq}
\end{center}
\end{figure}

A central question is connected to the conducting
properties of the resulting inhomogeneous state: Is it a global
superconductor, a metal, an insulator, or some unusual anisotropic
phase? To resolve this issue we use the concepts of charge stiffness
$D_c$ \cite{kohn}, which measures the sensitivity of the ground 
state to changes
in boundary conditions, and anomalous flux quantization \cite{fq}, which
provides a direct signature of the Meissner effect. These concepts,
together with topological quantum numbers \cite{gerardo1}, are 
routinely employed to study the localization properties of models
of interacting electrons. To compute $D_c$ one needs to determine how
the energy of a system with a fixed number of particles, $E$, depends on
the twist in the boundary conditions, $\Theta \in [0,2 \pi)$. The
twist of the boundary conditions is independently applied along each
spatial direction $j=x,y$ and implies that $c_{N_j+1} = \exp(i\Theta)
c_1$.  The special case of $\Theta = 0$ corresponds to strictly
periodic boundary conditions. Textbook schematics of the energy dependence,
$E(\Theta)$, are shown in Fig. \ref{fig:cq}A.  
The calculated many-particle spectrum $E(\Theta)$ for a system with a stripe
separation of 17 lattice sites in our model 
is shown in Fig. \ref{fig:cq}B. 
The energy curves imply that the system is
superconducting, however the superfluid stiffnesses along and across 
the stripes are drastically different.
However, for a smaller stripe periodicity (Fig. \ref{fig:cq}C), 
due to substantial 
overlap between the stripes, superconductivity is almost as strong across the
stripes as it is along the stripes.

Arrays of superconductors separated by insulating regions, known as
Josephson junction arrays, have non-trivial conducting properties.
Depending on the relative strength of the coupling between the
superconductors and the charging energy, 
such systems can be either superconductors or
insulators \cite{larkin}. At the MF level, the charging aspect of the problem
is absent, and hence we can expect a global superconductor at zero temperature
no matter how weak the interchain coupling is.  In a real system, for a 
sufficiently weak
coupling the superconducting behavior will be suppressed.
In the case of superconducting channels separated 
by AF insulators, the coupling is inversely related to
$L(x)$. For very large $L(x)$ at low doping, 
$L(x) \gg \xi_{SC}$,
the overlap between the superconducting stripes, 
and hence the superconducting transition temperature $T_c$,
is exponentially small.  
As the distance between the stripes decreases (larger doping),
the overlap of the superconducting condensate wave functions should
establish a phase coherent superconducting state.
Indeed, this is qualitatively what we observe already at the mean-field
level in the striped superconductors.
For a large superconducting stripe overlap, the effective coupling 
between the stripes is non-exponential.  In this regime, the experimentally
measured superconducting transition temperature is proportional to 
incommensuration \cite{balatsky}, 
which implies that the effective Josephson coupling scales as $1/L(x)$.

A possible experimental test of the Josephson-coupled superconductor
scenario proposed here (see also \cite{eroles}) can be performed by 
measuring the in-plane
Josephson plasmon resonance.  The resonance should be present in the 
microwave-frequency range and is excitable by an in-plane electric
field.  

\begin{figure}[htbp]
  \begin{center}
   \includegraphics[width = 3.0 in]{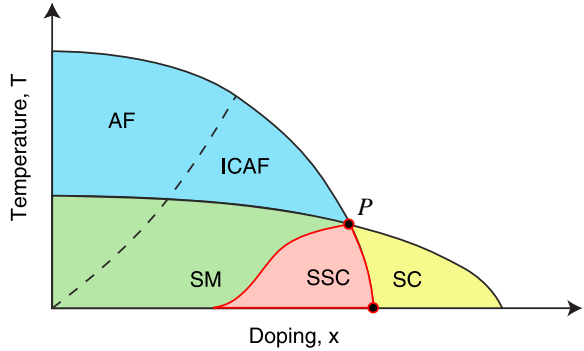}
\vspace{0.5cm}

\caption{Schematic phase diagram obtained by superimposing the 
antiferromagnetic (AF) / striped (ICAF) 
and the $d$-wave superconducting
(SC) phase diagrams. In the intersection region we distinguish the
subregions of  Josephson-coupled striped superconductor (SSC), and
non-superconducting ``strange metal'' 
(SM),  which is neither a superconductor, nor
a simple insulator.  The upper boundary  of the AF/ICAF corresponds to
the weak pseudogap crossover, and the line  between the pure AF/ICAF
and the SM marks the strong pseudogap  crossover.  In the 
very low doping region, the superconducting aspect of the problem becomes
irrelevant, and the phase diagram is  dominated by the physics
of antiferromagnets.  A detailed finite-temperature study is required to
precisely locate the left boundary of the SSC region, and hence to determine if
the critical point $P$ is indeed a penta-critical point.}

\label{fig:pd}
\end{center}
\end{figure}

From our zero-temperature analysis of the coexistence of AF stripes
(ICAF) and superconductivity, a simple qualitative thermodynamic phase
diagram emerges. 
In the conjectured phase diagram, we utilize the finite-temperature
AF/ICAF phase diagram of the Hubbard model
together with the superconducting (SC) phase diagram of the $t$-$V$ model.
The SC phase diagram is obtained in the homogeneous MF \cite{micnas}, 
while the AF/ICAF phase 
boundary is constructed under the assumption of the second order phase
transition between
the homogeneous and inhomogeneous states \cite{schulz}.
For a suitable choice of parameters, for instance $U = 2t$ 
and $V = - t$, the 
SC and the AF/ICAF regions in the phase diagram intersect, as shown in
Fig. \ref{fig:pd}. 
The boundary between the AF and ICAF phases corresponds to an infinite period
stripe modulation, and implies that the incommensuration is a 
decreasing function of temperature.
The energy scale associated with the AF/ICAF region
of the phase diagram is much larger than that of the SC part. Thus, one
expects that only the SC phase boundary is modified when it passes
through the AF/ICAF region. The central result of our work is that the
superconductivity {\em does not} disappear in the region of
the AF stripes, 
but rather becomes striped, with anisotropic superfluid stiffness.

Based on familiar Josephson coupling physics, in the region of
coexistence of superconductivity and stripes, we can expect a part
that is a globally coherent striped superconductor (SSC). The rest of
the intersection region is covered by an exotic phase  which, if it
were perfectly orientationally ordered, would be a superconductor in
one direction and a strongly-correlated insulator in the other. In
reality, due to the meandering of the stripes and their break-up into
finite segments \cite{fradkin}, the state is likely to be 
highly inhomogeneous and
neither an
insulator, nor a superconductor, but also not a simple metal. In
agreement with the experimental attribution, 
we refer to this region as
a ``strange metal'' (SM). The line separating the SM from
the AF/ICAF region, in the context of the experiments, can be
associated with the crossover to the strong pseudogap regime, and
corresponds to the opening of the superconducting gap. The
high-temperature AF/ICAF phase boundary marks to onset of the weak
pseudogap. For very small dopings the MF stripe separation becomes so
large that the superconducting aspects of the model become irrelevant
and one crosses over to the regime governed predominantly by the 
physics of antiferromagnets. 

It should be emphasized that the phase diagram presented here 
is based on the  (inhomogeneous) MF treatment of a 2-dimensional model.
As such, it is susceptible to the quantum and thermal 
fluctuations that tend to destroy long-range order.
For instance, the ICAF phase in Figure \ref{fig:pd}
in a real material is more likely to manifest itself as incommensurate AF
fluctuations, rather than a pure phase.  However, the effects of the
3rd dimension and impurity pinning may stabilize the MF phases at
a sufficiently low temperatures, revalidating the MF phase diagram.  

Within our model, we find that increasing on-site repulsion 
leads to a suppression of 
superconductivity.  For larger $U$, the pure $d$-wave superconducting region
of the phase diagram (SC) shrinks, and the width of the superconducting 
stripes in the SSC region decreases.  For example, changing $U$ from $4t$ to
$5t$ (keeping $V$ fixed) leads to a five-fold reduction of the superconducting
order parameter on stripes.
This is in contrast with the 
{\em homogeneous} mean-field results \cite{micnas}, 
there the d-wave superconductivity is independent of the magnitude of $U$.
On the other hand, large values of the on-site repulsion, $U > 4t$, 
in the Hubbard model lead to {\em diagonal} stripes \cite{schulz2}.
Applying our model to diagonal stripes ($U = 5t$, $V = -0.9t$), 
we find that superconductivity vanishes completely at the optimal 
stripe filling.
Perhaps it is not a coincidence, that the insulating cuprates do in fact show
diagonal stripes, as opposed to the lattice-aligned stripes in the
superconducting cuprates \cite{Birg}.

There is an analogy between the role played by doping in our model, and
the strength of the electron-phonon coupling, $\lambda$,
in the McMillan criterion for
the maximum achievable $T_c$ in conventional superconductors \cite{mcmill}.  
In McMillan's picture, increasing 
$\lambda$ favors superconductivity.  However, increasing $\lambda$ too much 
induces a structural transition, and hence changes the reference ground state.
In our model, increasing doping brings the superconducting stripes closer
together, and hence enhances the global $T_c$.  However, increasing doping 
too much causes a transition to the uniform state, with a subsequent 
monotonically decreasing $T_c$ as a function of doping.

The topology of the phase diagram we have proposed 
appears to be relevant
for the superconducting cuprates, such as LSCO and YBCO. The
same simple model (for other parameter values) can produce other 
topologies as well. A non-trivial
topology, which  may be realized in a material with a  weaker
attractive coupling, is when the superconducting region is fully 
contained in the AF/ICAF region.  Depending on the parameters, such a
material may be a striped superconductor in a certain range of dopings.
Also, similar  physics may occur in organic charge-transfer salts
\cite{brown}, that show AF, ICAF and superconductivity under pressure
which controls interchain coupling.  There, the intrinsically anisotropic 
coupling is also important for the stabilization of the mesoscopic 
inhomogeneities \cite{ivar}.

In conclusion, we find that a simple one-band model with  on-site 
repulsion and  nearest-neighbor attraction, in an appropriate
range of  parameters,
can simultaneously sustain both incommensurate antiferromagnetism  and
inhomogeneous superconductivity. Prompted by this finding and
utilizing  well-known antiferromagnetic and  superconducting phase
diagrams, we  have constructed a generic phase diagram that captures many
of the phases  observed in the cuprate and organic superconductors.
An experimental test of the Josephson-coupled superconductor
proposed here (see also \cite{eroles}) can be 
performed by measuring the in-plane Josephson plasmon resonance.
Although our simple model appears to capture much of the observed 
rich physics, however it 
can be readily elaborated (with multiple electron bands, 
3-dimensionality, long-range interactions, lattice coupling, etc.) for more
quantitative comparisons with specific materials.

We would like to thank  L. Bulaevskii for suggesting to us the 
in-plane Josephson plasma experiment as a probe of the striped 
superconductivity.  We acknowledge useful discussions with C.D. Batista, 
E. Fradkin, S. Kivelson, D. Morr, D. Pines, S. Trugman, and J. Zaanen. 
This work was supported by the U.S. DOE.


\end{multicols}

\begin{thebibliography}{99}
\bibitem{tranquada} 
J. M. Tranquada {\em et al.}, Phys. Rev. B {\bf 52}, 3581 (1995); 
M.~Arai {\em et al.}, Phys. Rev. Lett. {\bf 83}, 608 (1999); 
\mbox{Z.-X.~Shen {\em et al.}}, Science {\bf 280}, 259 (1998).
\bibitem{schulz}
H.J. Schulz, Phys. Rev. Lett. {\bf 64}, 1445 (1990).
\bibitem{zaanen}
J. Zaanen and O. Gunnarsson, Phys. Rev. B {\bf 40}, 7391 (1989).
\bibitem{poil}
D. Poilblanc and T. M. Rice, Phys. Rev. B {\bf 39}, 9749 (1989).
\bibitem{machida} K. Machida, Physica C {\bf 158} 192 (1989).
\bibitem{micnas} R. Micnas {\em et al.}, Rev. Mod. Phys. {\bf 62}, 113 (1990).
\bibitem{ivar2} I. Martin, unpublished.
\bibitem{ws} 
S. R. White and D. J. Scalapino, Phys. Rev. B {\bf 60}, R753 (1999). 
\bibitem{kohn} W. Kohn, Phys. Rev. {\bf 133}, A171 (1964). 
\bibitem{fq} N. Buyers and C. N. Yang, Phys. Rev. Lett. {\bf 7}, 46 (1961);
A. Sudbo {\em et al.}, Phys. Rev. Lett. {\bf 70}, 978 (1993).
\bibitem{gerardo1} 
G. Ortiz and R. M. Martin, Phys. Rev. B {\bf 49}, 14202 (1994).
\bibitem{larkin} G. Schon and A.D. Zaikin, Phys. Rep. {\bf 198}, 237 (1990). 
\bibitem{balatsky} 
A. V. Balatsky and Z.-X. Shen, Science {\bf 284}, 1137 (1999). 
\bibitem{eroles} J. Eroles {\em et al.}, to appear in Europhys. Lett (2000); 
preprint cond-mat/0003322 (2000).
\bibitem{fradkin} Kivelson, S.A. {\em et al.}, Nature {\bf 393}, 550 (1998); 
	B. Stojkovic {\em et al.}, Phys. Rev. Lett. {\bf 82}, 4679 (1999).
\bibitem{schulz2} H. J. Schulz, J. Phys. France {\bf 50}, 2833 (1989).
\bibitem{Birg} M. Matsuda {\em et al.}, preprint cond-mat/0003466 (2000).
\bibitem{mcmill} W. L. McMillan, Phys. Rev. {\bf 167}, 331 (1968).
\bibitem{brown} 
D. S. Chow {\em et al.}, Phys. Rev. Lett. {\bf 81}, 3984 (1998).
\bibitem{ivar} I. Martin {\em et al.}, unpublished.  
\end{thebibliography}
\end{document}